\documentclass[11pt]{article}
\usepackage{geometry}                
\usepackage{graphicx}
\usepackage{subfigure}
\usepackage{amssymb, amsmath}
\usepackage{epstopdf}
\newcommand{\BE}{\begin{equation}}
\newcommand{\EE}{\end{equation}}
\usepackage{float}
\title{Series extension: Predicting approximate series coefficients from a finite number of exact coefficients.}
\author{Anthony J Guttmann}
\date{23 April 2016}                                           

\begin{document}
\maketitle
\begin{abstract}
Given the first 20-100 coefficients of a typical generating function of the type that arises in many problems of statistical mechanics or enumerative combinatorics, we show that the method of differential approximants performs surprisingly well in predicting (approximately) subsequent coefficients. These can then be used by the ratio method to obtain improved estimates of critical parameters. In favourable cases, given only the first 20 coefficients, the next 100 coefficients are predicted with useful accuracy. More surprisingly, this is also the case when the method of differential approximants does not do a useful job in estimating the critical parameters, such as those cases in which one has stretched exponential asymptotic behaviour. Nevertheless, the coefficients are predicted with surprising accuracy. As one consequence, significant computer time can be saved in enumeration problems where several runs would normally be made, modulo different primes, and the coefficients constructed from their values modulo different primes. Another is in the checking of newly calculated coefficients. We believe that this concept of approximate series extension opens up a whole new chapter in the method of series analysis.
\end{abstract}

\section{Introduction}
Series analysis  has, for many years, been a powerful tool in the study of many problems in statistical mechanics, combinatorics, fluid mechanics and computer science. The name {\em series analysis} subsumes a range of numerical techniques designed to answer the following question: Given the first $N$ coefficients of the series expansion of some function, (where $N$ is typically as low as 5 or 6, or as high as 100,000 or more), determine the asymptotic form of the coefficients, subject to some underlying assumption about the asymptotic form, or, equivalently, the nature of the singularity of the function.

Typical examples include the susceptibility of the Ising model, and the generating function of self-avoiding walks (SAWs). These are believed to behave as 
\begin{equation}\label{generic}
F(z) = \sum_n c_n z^n \sim C \cdot (1 - z/z_c)^{-\gamma}.
\end{equation}
Here $z_c$ is the radius of convergence of the series expansion, and its reciprocal $\mu=1/z_c$ is known as the growth constant, as the dominant term in the asymptotics of the coefficient $c_n$ is $\mu^n.$
  In the Ising case, for regular two-dimensional lattices, the values of both $z_c$ and $\gamma=7/4$ are exactly known, and the amplitude $C$ is known to more than 100 decimal places. The value of $z_c$ depends on the choice of underlying lattice. In the SAW case, $z_c$ is only known for the hexagonal lattice \cite{DC10}, and the value  $\gamma=43/32$ is universally believed, but not proved. 

The method of series analysis is typically used when one or more of the critical parameters is not known. For example, for the three-dimensional versions of the above problems, none of the quantities $C,$ $z_c$ or $\gamma$ is exactly known.
 From the binomial theorem it follows from (\ref{generic}) that 
\begin{equation}\label{asymp1}
c_n \sim \frac{C }{\Gamma(\gamma)}\cdot  \mu^{n} \cdot  n^{\gamma-1}.
\end{equation}
 Here $C, \,\, \mu=1/z_c, \,\, {\rm and} \,\, \gamma$ are referred to as the critical amplitude, the growth constant ($z_c$ is the critical point) and the critical exponent respectively\footnote{Sometimes $ \frac{C }{\Gamma(\gamma)}$  is referred to as the critical amplitude}.

The aim of series analysis is to obtain, as accurately as possible, estimates of the critical parameters from the first $N$ coefficients. Since calculating these coefficients is typically a problem of exponential complexity, the usual consequence is that fewer than 100 terms are known (and in some cases far fewer)\footnote{In the case of the susceptibility of the two dimensional Ising model, polynomial time algorithms for enumerating the coefficients have been developed \cite{O00,C10}, and in that case we have hundreds, in some cases thousands, of terms. Unfortunately, this is a rare situation.}.

There are literally thousands of situations in statistical mechanics, combinatorics, computer science and fluid mechanics (and other areas) where such problems arise. 

The methods used to extract estimates of the critical parameters from the known expansion coefficients largely fall into two classes. One class is based on the {\em  ratio method}, initially developed by Domb and Sykes \cite{DS56}, and subsequently refined and extended by many authors. A related, but not fully equivalent idea is that of direct fitting to coefficients or ratios. This requires rather precise knowledge of the underlying asymptotics, as it involves fitting to several sub-dominant terms. A typical example of this approach can be found in \cite{CG05}.

The second is based on analysing a linear ordinary differential equation (ODE), the solution of which  has an algebraic singularity (\ref{generic}). The ODE (in fact there are usually many ODEs using a given number of series coefficients) is constructed so that the first $N$ terms of the power series expansion of its solution precisely agrees with the known series expansion coefficients. The first development of this nature was due to Baker \cite{GAB61}, based on taking Pad\'e approximants of the logarithmic derivative of known series, and is described in section \ref{ana:pade} below. The point is that if one constructs an $[L/M]$  Pad\'e approximant to the logarithmic derivative of a function $f(z),$ this corresponds to fitting the first $L+M+1$ coefficients of the series expansion of $f(z)$ to the linear ODE $P_L(z)f'(z)-Q_M(z)f(z)=0,$ with $P_L(z)$ and $Q_M(z)$ being polynomials of degree $L$ and $M$ respectively, (for uniqueness, one usually chooses $Q_M(0)=1$). This idea was then substantially extended by Guttmann and Joyce \cite{GJ72} who developed the {\em method of differential approximants}, which is still the most successful method in use today for analysing series with algebraic singularities, typified by (\ref{generic}).

Though many problems have such an algebraic singularity structure, an increasing number of situations have been encountered \cite{G14,BGJL14} in which a more complex structure prevails. Those cases are characterised by the presence of an additional {\em stretched exponential} term in the asymptotic form of the coefficients,  which then behave as
\begin{equation}\label{asymp2}
b_n \sim C \cdot z_c^{-n}\cdot \mu_1^{n^\sigma}\cdot n^{g}.
\end{equation}
That is to say, there is a sub-dominant stretched exponential term $\mu_1^{n^\sigma},$ giving rise to two additional parameters, $\mu_1$ and $\sigma.$ If $\mu_1 > 1,$ this term dominates the term $n^g$ carrying the critical exponent, and one can write down a generic generating function whose coefficients have this asymptotic behaviour \cite{FS09}. But if, as is usually the case, $\mu_1 < 1,$ then the stretched exponential term is eventually dominated by the term $n^g,$ and furthermore a generic generating function does not appear to be known in this case. 

It was shown in \cite{G14} that both the ratio method and the method of differential approximants need to be modified if reliable estimates of even some of the critical parameters are to be obtained in this more complex case.

In the next sections we give a brief review of the two methods, and then go on to discuss the major point of this paper, which is that the two methods work surprisingly well together in (approximately) extending the number of known series coefficients. More precisely, the differential approximants (DAs) predict, in principle, all coefficients beyond those used to construct the approximant. The accuracy of the predicted coefficients decreases with increasing order of the coefficients, but in many, indeed most cases, a substantial number of approximate coefficients can be obtained with sufficient accuracy as to be useful in applying the ratio method.

For an isolated algebraic singularity (\ref{generic}), with no nearby competing singularities, the method of DAs is so effective that the ratio method cannot compete in terms of precision of estimates of the critical point or exponent. However in more difficult situations, such as in the presence of competing singularities, or more particularly, stretched exponential behaviour, the insight given by the ratio method with the (approximate) predicted terms is extremely valuable.

As a separate advantage, recall that the exact series coefficients are frequently hard won, as the underlying counting problem is usually of exponential complexity, sometimes requiring hundreds of hours of computer time for the last coefficient. The fact that one can often predict these coefficients to 15-20 digit precision in less than a second is remarkable. It is also useful in saving computer time, as we explain in section \ref{sec:cpu}. A second advantage is in checking newly computed coefficients. It occasionally happens that the last predicted coefficient of a series is slightly wrong. By predicting the last coefficients from all previous coefficients, such errors can sometimes be detected.

\section{Ratio Method}\label{Ratio}
The ratio method \index{ratio method} was perhaps the earliest systematic
method of series analysis employed,
and is still a useful starting point, prior to the application of more sophisticated
methods. It was first used by M~F~Sykes in his 1951 D Phil studies, under the supervision
of C~Domb. From equation (\ref{asymp1}),
it follows that the {\it ratio} of successive terms is
\begin{equation} \label{ratios}
r_n = \frac{c_n}{c_{n-1}}=\frac{1}{z_c}\left (1 + \frac{\gamma -1}{n} + {\rm o}\left (\frac{1}{n}\right )\right ).
\end{equation}
 From this equation, it is natural to plot the successive ratios $r_n$ against $1/n.$
If the correction terms ${\rm o}\left (\frac{1}{n}\right )$ can be ignored\footnote{For a purely algebraic singularity (\ref{generic}), with no confluent terms, the correction term will be ${\rm O}\left (\frac{1}{n^2}\right )$ in (\ref{ratios}).}, such a plot will be linear,
with gradient $\frac{\gamma-1}{z_c},$ and intercept $1/z_c$ at $1/n = 0.$

\begin{figure}
\centerline {
\includegraphics[width=10cm]{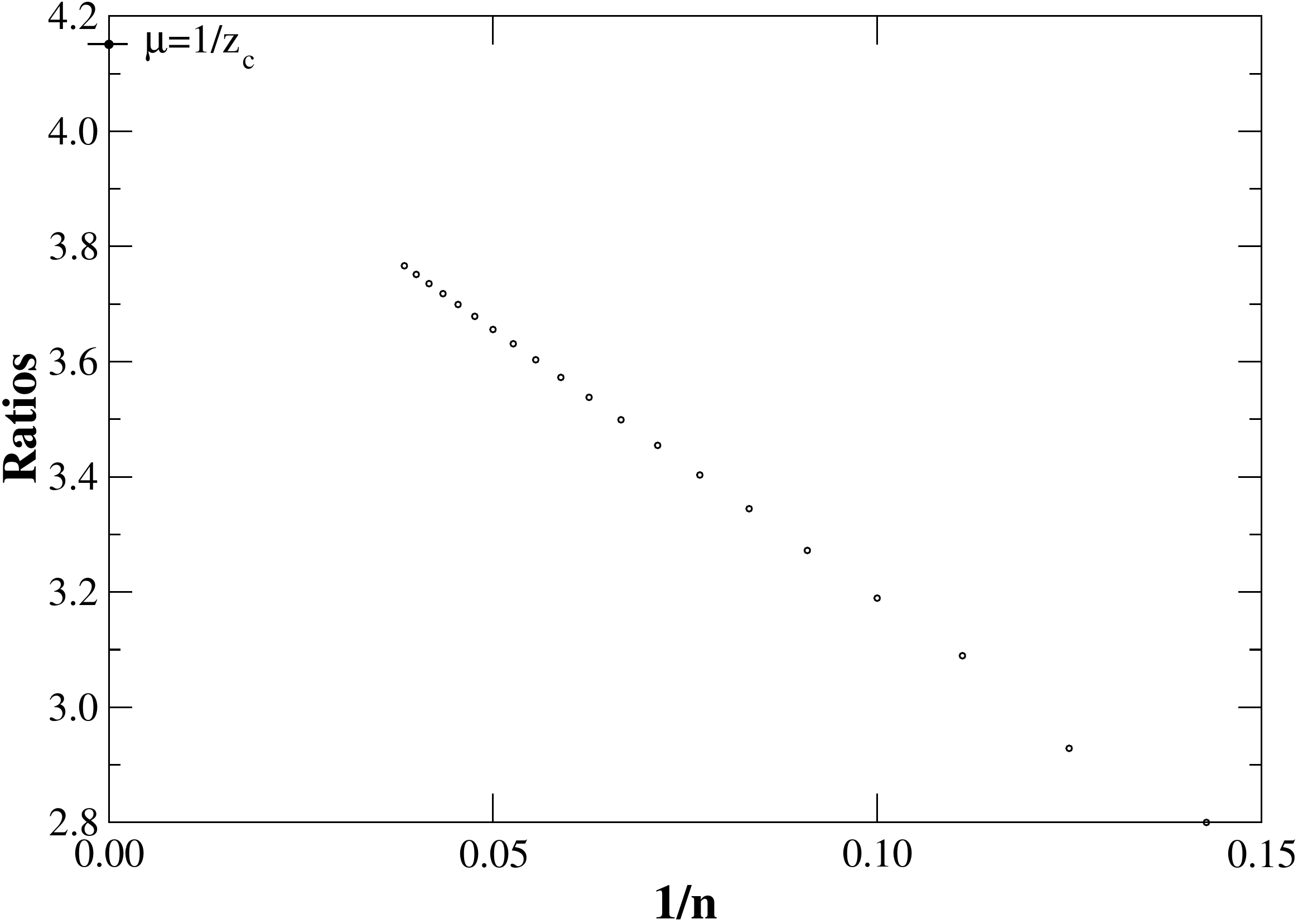}
}
\caption{\label{fig:ratt}
Plot of ratios against $1/n$ for triangular lattice polygons. A straight line through the
last few data points intercepts the Ratios axis approximately at $1/z_c.$
}
\end{figure}

As an example, we apply the ratio method to the generating function of self-avoiding polygons (SAPs) on the triangular lattice. The first few terms in
the generating function, from $p_3$ to $p_{26}$ are:
 2,
    3,
   6,
    15,
   42,
    123,
    380,
   1212,    3966,   13265,
   45144,
   155955,
   545690,
   1930635,
   6897210,
   24852576,
   90237582,
  329896569,
   1213528736,
   4489041219,
  16690581534,
  62346895571,
   233893503330,
   880918093866.
Plotting successive ratios against $1/n$ results in the plot shown in Figure \ref{fig:ratt}.
The critical point is estimated \cite{Jensen04} to be at $z_c \approx 0.240917574\ldots = 1/4.15079722\ldots .$
   \vspace{-0.2in}
\vspace{.2in}

From the figure one sees that the locus of points, after some initial (low $n$) curvature,
becomes linear to the naked eye for $n > 15$ or so, (corresponding to $1/n < 0.067$).
Visual extrapolation to $1/z_c$ is quite obvious. A straight line drawn through the
last $4-6$ data points intercepts the horizontal axis around $1/n \approx 0.13.$ Thus
the gradient is approximately $\frac{4.1508-2.8}{-0.13} \approx -10.39,$
from which we conclude that the exponent $\gamma - 1 \approx -10.39 \cdot z_c \approx -2.50.$ It is believed \cite{N82}
that the exact value is $\gamma = -3/2,$ which is in complete agreement with this
simple graphical analysis.

Various refinements of the method can be readily derived. From (\ref{ratios}) it follows that estimators of the growth constant $\mu = 1/z_c$  are given by
\begin{equation}
\label{muest}
\mu_n = nr_n - (n-1)r_{n-1} = \mu \left(1 + o\left( \frac{1}{n}  \right )\right ).
\end{equation}
If the critical point
is known exactly, it follows from equation (\ref{ratios}) that estimators of the exponent
$\gamma$ are given by
\begin{equation} \label{gamest}
 \gamma_n=n(z_c\cdot r_n-1)+1 = \gamma +  {\rm o}(1).
 \end{equation}

If $z_c$ is unknown, estimates of the exponent $\gamma$ can be obtained by defining estimators $\gamma_n$ of $\gamma$ and extrapolating these against $1/n.$ Here
\BE \label{eq:exp}
\gamma_n=1+n^2\left ( 1-\frac{r_n}{r_{n-1}} \right )= \gamma +  {\rm o}(1).
\EE
Similarly, if only the exponent $\gamma$ is known, estimators of the critical point $z_c$
are given by  $$z_c^{(n)} =  \frac{n+\gamma-1}{n r_n}= z_c + o\left( \frac{1}{n}  \right ).$$
 In all the above cases, if the singularity is a simple algebraic singularity, with no  confluent terms, the correction term $o\left (1 \right )$ can be replaced by $O\left ( \frac{1}{n} \right ),$ and $o\left( \frac{1}{n}  \right )$ by $O\left( \frac{1}{n^2}  \right ).$
 
 Finally, the amplitude $C$ is seen from (\ref{asymp1}) to be estimated by extrapolating the sequence 
 \begin{equation}\label{eq:amp}
 C_n =c_n\Gamma(\gamma)z_c^n n^{1-\gamma} \sim C + o(1)
 \end{equation}
  against $1/n.$

One problem with the ratio method is that if the singularity closest to the origin
is not the singularity of interest (the so-called {\it physical singularity}),
\index{physical singularity} then
the ratio method will not give information about the physical singularity. Worse still,
if the closest singularity to the origin is a conjugate pair, the ratios will vary dramatically in both
sign and magnitude. To overcome this difficulty G A Baker Jr \cite{GAB61} proposed
the use of Pad\'e approximants applied to the logarithmic derivative of the series expansion, as mentioned above and discussed in section \ref{ana:pade} below.

We should also mention that there exists a vast literature of extrapolation techniques in numerical analysis, and many such methods can be advantageously applied to extrapolate the sequence of ratios in order to estimate the radius of convergence, which is the critical point. Some of these methods, applied to series analysis problems, are discussed in the review \cite{G89}. 

\section{Pad\'e approximants} 
\label{ana:pade}

The basic idea of using Pad\'e approximants in series analysis is very simple. For a meromorphic function $F(z)$ we use its series expansion 
to form a rational approximation to $F(z),$ 

\BE \label{eq:ana_pade}
F(z) = \frac{P_i(z)}{Q_j(z)}
\EE
where $P_i(z)$ and $Q_j(z)$ are polynomials of degree $i$ and $j$ respectively,
whose coefficients are chosen such that the first $i+j+1$ terms in the series expansion
of $F(z)$ are identical to those of the expansion of $P_i (z)/Q_j(z),$ with $Q_j(0)=1$ for uniqueness. Constructing the polynomials only involves solving a system of linear equations.

In order to use Pad\'e approximants  to reliably approximate an algebraic singularity rather than just
poles, 
we must first transform the series into a suitable form. 
If we have a function with  expected behaviour typical of algebraic singular
points, as given
by (\ref{generic}),
then taking the derivative of the logarithm of $F(z)$ gives
\BE \label{eq:ana_Dlog}
\widehat{F}(z)=\frac{\rm d}{{\rm d}z} \log F(z) \simeq \frac{\gamma}{z_c-z} +{\rm o}\left (\frac{1}{z_c-z}\right ).
\EE
This form is perfectly suited to Pad\'e analysis, as taking the logarithmic derivative has turned the function into a meromorphic function (at least to leading order). 
Estimates of the critical point $z_c$ can be obtained from the
roots of the denominator polynomial $Q_j(z)$, while estimates of the
critical exponent $\gamma$ are
obtainable from the residue of the Pad\'e approximant to $\widehat{F}(z)$ at $z_c$,
that is
\BE \label{eq:ana_resexp}
\gamma \approx \lim_{z\to z_c} (z_c-z)\frac{P_i(z)}{Q_j(z)}.
\EE
As is the case for the ratio method, refinements exist for those situations when the critical point or critical exponent is exactly known \cite{G89}.

Since $\widehat{F}(z) = F'(z)/F(z),$ we see that
forming a Dlog-Pad\'e approximant is simply equivalent to seeking an approximation
to $F(z)$ by solving the first order homogeneous differential equation

$$F'(z)Q_j(z)-F(z)P_i(z) =0.$$
This observation leads us directly to the more powerful and more general method
of {\em differential approximants} by noting that one
can approximate $F(z)$ by a solution to a higher order ODE (possibly inhomogeneous).
 This method was first proposed and developed by
Guttmann and Joyce \cite{GJ72} in 1972, and was subsequently extended to the
inhomogeneous case by Au-Yang and Fisher \cite{FA79} and Hunter and Baker \cite{HB79} in 1979. It is not uncommon for the problems under consideration to have a more complex singularity structure, wherein there is more than one singularity at the radius of convergence. Typically one then has behaviour of the form
\begin{equation}
\label{confluent}
F(z) = \sum_n c_n z^n \sim C_1 \cdot (1 - z/z_c)^{-\gamma} + C_2 \cdot (1 - z/z_c)^{-\gamma+ \Delta},
\end{equation}
where the {\em confluent exponent} $\Delta$ is not a positive integer. We refer to such a singularity as {\em confluent}.
The advantage of a higher order ODE is that confluent singularities can be accommodated \cite{RJG80}, as well as a more complicated singularity structure in general. Functions that satisfy such an ODE are called {\em D-finite} or {\em holonomic}.

\section{Differential approximants}
\label{ana:da}

The generating
functions  of  lattice models
in statistical mechanics and enumerative combinatorics are sometimes algebraic, such as that for Dyck paths, or the magnetisation of the two-dimensional Ising model, or sometimes D-finite such as the internal energy of the two-dimensional Ising model.
This latter observation is the origin of the method of {\em differential approximants}.
The basic idea is to approximate a generating function $F(z)$ by solutions
of differential equations with polynomial coefficients. The singular behaviour
of such ODEs is  well documented
(see e.g. \cite{Forsyth02,Ince27}), and the singular points and
exponents are readily calculated from the ODE. The key point for series analysis is that even if {\em globally} the function is not describable by a solution
of such a linear ODE (as is frequently the case) one expects that
{\em locally,} in the
vicinity of the (physical) critical points, the generating
function is still well-approximated by a solution of a linear ODE, when the singularity is of generic algebraic type (\ref{generic}).

An $M^{th}$-order differential approximant (DA) to a function $F(z)$  is formed by matching
the coefficients in the polynomials $Q_k(z)$ and $P(z)$ of degree $N_k$ and $L$, respectively,
so that the formal solution of the $M^{th}$-order inhomogeneous ordinary differential equation
\BE \label{eq:ana_DA}
\sum_{k=0}^M Q_{k}(z)(z\frac{{\rm d}}{{\rm d}z})^k \tilde{F}(z) = P(z)
\EE
agrees with the first $N=L+\sum_k (N_k+1)$ series coefficients of $F(z)$. Constructing such ODEs only involves
solving systems of linear equations. The function
$\tilde{F}(z)$ thus agrees with the power series expansion of the (generally unknown)
function $F(z)$ up to the first $N$ series expansion coefficients.
We normalise the DA by setting $Q_M(0)=1,$ thus leaving us with $N$ rather
than $N+1$ unknown coefficients to find. The choice of the differential operator $z\frac{{\rm d}}{{\rm d}z}$ in (\ref{eq:ana_DA}) forces the origin to be a regular singular point. The reason for this choice is that most lattice models with holonomic solutions, for example, the free-energy of the two-dimensional Ising model, possess this property. However this is not an essential choice.

Our notation for a differential approximant is $$[deg(Q_M),deg(Q_{M-1}), \cdots , deg(Q_0),deg(P)],$$ where $deg(Q_T)$ denotes the degree of polynomial $Q_T(z)$ in equation (\ref{eq:ana_DA}).

From the theory of ODEs, the singularities of $\tilde{F}(z)$ are approximated by zeros
$z_i, \,\, i=1, \ldots , N_M$ of $Q_M(z),$ and the
associated critical exponents $\gamma_i$ are estimated from the indicial equation. If there is only a single root at $z_i$  this is just
\BE \label{eq:ana_indeq1}
\gamma_i=M-1-\frac{Q_{M-1}(z_i)}{z_iQ_M ' (z_i)}.
\EE
Estimates of the critical amplitude $C$ are rather more difficult to make, involving the integration of the differential approximant. For that reason the simple ratio method approach to estimating critical amplitudes is often used, whenever possible taking into account higher-order asymptotic terms \cite{GJ09}.

Details as to which approximants should be used and how the estimates from many approximants are averaged to give a single estimate are given in \cite{GJ09}. Examples of the application of the method can be found in \cite{G14}. In that work, and in this, we reject so-called {\em defective} approximants, typically those that have a spurious singularity closer to the origin than the radius of convergence as estimated from the bulk of the approximants. Another  method sometimes used is to reject outlying approximants, as judged from a histogram of the location of the critical point (i.e. the radius of convergence) given by the DAs. It is usually the case that such distributions are bell-shaped and rather symmetrical, so rejecting approximants beyond two or three standard deviations is a fairly natural thing to do.

\section{Coefficient prediction}
\label{pred}
In this paper we show that the ratio method and the method of differential approximants  work serendipitously together in many cases, even in the situation where one has stretched exponential behaviour, in which case neither method works particularly well in unmodified form. To be more precise, the method of differential approximants (DAs)  produces ODEs which, by construction, have solutions whose series expansions agree term by term with the known coefficients used in their construction. Clearly, such ODEs implicitly define {\em all}  coefficients in the generating function, but if $N$ terms are used in the construction of the ODE, all terms of order $z^{N+1}$ and beyond will be approximate, unless the exact ODE is discovered, in which case the problem is solved, and we have no need to recourse to approximate methods.

What we have found is that it is useful is to construct a number of DAs that use all available coefficients, and then use these to predict subsequent coefficients. Not surprisingly, if this is done for a large number of approximants, it is found that the predicted coefficients of the term $z^n,$ where $n > N,$ agree for the first $k(n)$ digits, where $k$ is a decreasing function of $n.$ We take as the predicted coefficients the mean of those produced by the various DAs, with outliers excluded, and as a measure of accuracy we take the number of digits for which the predicted coefficients agree, or the standard deviation. These two measures of uncertainty are usually in good agreement.

Now it makes no logical sense to use the approximate coefficients as input to the method of differential approximants, as we have used the DAs to obtain these coefficients. However there is no logical objection to using the ({\em approximate}) predicted coefficients as input to the ratio method. Indeed, as the ratio method, in its most primitive form, looks at a graphical plot of the ratios, an accuracy of 1 part in $10^3$ or $10^4$ is sufficient, as errors of this magnitude are graphically unobservable. 

Recall that, in the ratio method one looks at {\em ratios} of successive coefficients. We find that the ratios of the approximate coefficients are predicted with even greater precision than the coefficients themselves by the method of DAs. That is to say, while a particular coefficient and its successor might be predicted with an accuracy of 1 part in $10^p$ for some value of $p$, the {\em ratio} of these successive coefficients is frequently  predicted with significantly greater accuracy (the precision being typically improved by a factor varying between 2 and 20). In favourable cases, this idea is surprisingly effective.

This technique was recently used moderately effectively for the study of bridges and terminally attached SAWs on the simple-cubic lattice in \cite{CCG15}. To be precise, the next 7 coefficients were predicted from a number of DAs, and then used in a ratio analysis, which revealed behaviour not obvious from the ratios of the (exact) available series. As we show below, the authors were probably too tentative predicting only the next seven coefficients.

We only have a qualitative understanding of why this idea of series extension, as described, is so effective. The DAs use all the information in the coefficients, and are sensitive to even quite small errors in the coefficients. As an example, in a recent study of some self-avoiding walk series, an error was detected in the twentieth significant digit in a new coefficient, as the DAs were much better converged without the last, new, coefficient. They also require high numerical precision in their calculation. In favourable circumstances, they can give remarkably precise estimates of critical points and critical exponents, by which we mean up to or even beyond 20 significant digits in some cases. Remarkably, this can be the case even when the underlying ODE is not D-finite. Of course, the singularity must be of the assumed algebraic form.

Ratio methods, and direct fitting methods, by contrast are much more robust. The sort of small error that affects the convergence of DAs would not affect the behaviour of the ratios, or their extrapolants, and would thus be invisible to them. As a consequence, approximate coefficients are just as good as the correct coefficients in such applications, provided they are accurate enough. We re-emphasise that, in the generic situation (\ref{generic}), ratio type methods will never give the level of precision in estimating critical parameters that DAs can give. By contrast, the behaviour of ratios can more clearly reveal features of the asymptotics, such as the fact that a singularity is not algebraic. This is revealed, for example, by curvature of the ratio plots \cite{G14}.

A related observation\footnote{I would like to thank my colleague Nathan Clisby for pointing this out.} is that the ordinary Pad\'e approximants, which of course can only accurately represent meromorphic functions, can also be used to predict coefficients of any given generating function. When applied to OGFs with known algebraic singularities, while they give very poorly converged estimates of the location of the radius of convergence, and even worse estimates of the value of the critical exponent, they can predict subsequent coefficients with some accuracy.

\subsection{Two-dimensional Ising susceptibility series}
As our first example, in which admittedly the method is seen to its best advantage, we consider the high-temperature susceptibility series of the Ising model on the triangular lattice. Several hundred series coefficients are known \cite{C10}, but let us assume we only know the first twenty coefficients, which is often the case with more difficult problems. We have used these twenty coefficients to predict the subsequent 100 coefficients! In table \ref{isingt20} we show the predicted value of the 120th and 121st coefficient, $c_{120}$ and $c_{121}$ respectively, as predicted by second-order differential approximants, and in subsequent columns we show the error in these predicted coefficients. In further columns we show the predicted ratio $c_{121}/c_{120}$ and the associated error. Note that the coefficients are predicted with an error typically of a few parts in $10^5,$ while the ratios are predicted with an accuracy some 50 times greater. So the error in the ratios is a few parts in $10^7$ in most cases, or a few parts in $10^6$ in the worst cases. This is more than enough precision for the ratio method to be very effectively used.

In table  \ref{isingt20a} we show corresponding results obtained from 3rd order DAs, and it can be seen that the results are comparable. In the next two tables we give the results when we use forty terms of the susceptibility series instead of 20, and use these to predict the next two hundred coefficients. In tables \ref{isingt40} and \ref{isingt40a} we give the predicted value of the 240th and 241st coefficient, $c_{240}$ and $c_{241}$ respectively, as predicted by third-order differential approximants and fourth-order approximants respectively. Note that the coefficients are predicted with an error typically of a few parts in $10^9,$ while the ratios are predicted with an accuracy some 50 times greater. So the error in the ratios is a few parts in $10^{11}$ in most cases, or a few parts in $10^{12}$ in the best cases. This is astonishing precision. It is remarkable that with 40 coefficients one can predict the first 9 or more digits of the next 200 coefficients. The fourth-order approximants do not do as well, predicting about 1 fewer digit in the coefficients, and with the ratios only being about twice as precise as the coefficients.

Next we turn to the high-temperature susceptibility series of the square lattice Ising model. This series has two singularities on the circle of convergence, one at $z=z_c$ corresponding to the ferromagnetic critical point, the other at $z=-z_c,$ corresponding to the anti-ferromagnetic critical point. This makes the asymptotics a bit more complicated, being the sum of two terms, one of which is of constant sign, the other is of alternating sign.  One would expect the DAs to struggle a bit more to successfully simulate this function, and we see this in the (comparatively) reduced precision of predicted coefficients. 

Once again we have used just twenty coefficients to predict the subsequent 100 coefficients. In table \ref{isings20} we show the predicted value of the 120th and 121st coefficient, $c_{120}$ and $c_{121}$ respectively, as given by both second-order and third-order differential approximants, taking as input only the coefficients up to $c_{20}.$ Note that the coefficients and ratios are predicted with an error typically of a few parts in $10^2,$ or $10^3.$ Clearly seeking 100 further terms from just 20 terms is over-ambitious. In the next table, table \ref{isings40}. we show the corresponding results, this time using the first 40 terms in the series to predict the next 100 terms. This is rather more successful, with both the coefficients and the ratios being predicted with an accuracy of a few parts in $10^6$ or $10^7.$ This is more than adequate to be useful in a ratio analysis.

\begin{table}[!ht]
\caption{Prediction of $c_{120} $ and $c_{121}$ from triangular Ising susceptibility series to order $z^{20}$ by second order inhomogeneous DAs. The ratio $c_{121}/c_{120}$ is calculated, and errors  are given. Note that the error in the ratios is typically less than $1/50$th the error in the coefficients. Here and in subsequent tables $e$ refers to a power of $10$, so that $e69, \,\, e-5$ means $10^{69}$ and  $ 10^{-5}$ respectively.}
\begin{center}
 \scalebox{0.9}{
\begin{tabular}{|ccccccc|}
\hline \\
Approximant & $c_{120} $ & Error & $c_{121}$ & Error & Ratio & Error \\
\hline
Exact & $3.85433525\ldots e69$ & 0 & $1.44743231\ldots e70$ & 0 & $3.75533579\ldots $&0 \\
$[5,6,6,1]$ & $ 3.85409947e69$ & $6.1e-5$ & $1.44734213e70$& $6.2e-5$ & $3.7553315$ & $1.2e-6$ \\
$[6,5,6,1]$ & $ 3.85429207e69$ & $1.0e-5$ & $1.44741580e70$& $1.1e-5$ & $3.7553350$ & $2.1e-7$ \\
$[6,6,5,1]$ & $ 3.85425583e69$ & $2.0e-5$ & $1.44740194e70$& $2.1e-5$ & $3.7553344$ & $3.7e-7$ \\
$[5,5,6,2]$ & $ 3.85416725e69$ & $4.3e-5$ & $1.44736801e70$& $4.4e-5$ & $3.7553327$ & $8.3e-7$ \\
$[5,6,5,2]$ & $ 3.85414356e69$ & $5.0e-5$ & $1.44735897e70$& $5.1e-5$ & $3.7553323$ & $9.3e-7$ \\
$[6,5,5,2]$ & $ 3.85438659e69$ & $1.3e-5$ & $1.44745199e70$& $1.4e-5$ & $3.7553368$ & $2.7e-7$ \\
$[5,5,5,3]$ & $ 3.85417010e69$ & $4.3e-5$ & $1.44736912e70$& $4.4e-5$ & $3.7553328$ & $8.0e-7$ \\
$[5,6,4,3]$ & $ 3.85419657e69$ & $3.6e-5$ & $1.44737935e70$& $3.7e-5$ & $3.7553332$ & $6.9e-7$ \\
$[4,5,5,4]$ & $ 3.85416591e69$ & $4.6e-5$ & $1.44736752e70$& $4.7e-5$ & $3.7553327$ & $8.3e-7$ \\
$[5,4,5,4]$ & $ 3.85415916e69$ & $4.6e-5$ & $1.44736494e70$& $4.7e-5$ & $3.7553326$ & $8.5e-7$ \\
$[5,5,4,4]$ & $ 3.85419118e69$ & $3.7e-5$ & $1.44737715e70$& $3.8e-5$ & $3.7553330$ & $7.5e-7$ \\
$[4,4,5,5]$ & $ 3.85414063e69$ & $5.1e-5$ & $1.44735784e70$& $5.2e-5$ & $3.7553322$ & $9.6e-7$ \\
$[4,5,4,5]$ & $ 3.85423988e69$ & $2.5e-5$ & $1.44739583e70$& $2.5e-5$ & $3.7553341$ & $4.5e-7$ \\
$[5,4,4,5]$ & $ 3.85447693e69$ & $3.8e-5$ & $1.44748654e70$& $3.8e-5$ & $3.7553384$ & $6.9e-7$ \\
$[4,3,5,6]$ & $ 3.85394654e69$ & $1.0e-4$ & $1.44728342e70$& $1.0e-4$ & $3.7553283$ & $2.0e-6$ \\
$[4,4,4,6]$ & $ 3.85431446e69$ & $5.5e-6$ & $1.44742439e70$& $5.5e-6$ & $3.7553355$ & $8.0e-8$ \\
$[3,4,4,7]$ & $ 3.85384669e69$ & $1.3e-4$ & $1.44724547e70$& $1.3e-4$ & $3.7553271$ & $2.3e-6$ \\
$[4,3,4,7]$ & $ 3.85372780e69$ & $1.6e-4$ & $1.44720070e70$& $1.6e-4$ & $3.7553266$ & $2.4e-6$ \\
$[3,3,4,8]$ & $ 3.85370226e69$ & $1.6e-4$ & $1.44719025e70$& $1.7e-4$ & $3.7553245$ & $3.0e-6$ \\
\hline

\end{tabular}
}
\end{center}
\label{isingt20}
\end{table}%

 \begin{table}[H]
\caption{Prediction of $c_{120} $ and $c_{121}$ from triangular Ising susceptibility series to order $z^{20}$ by third order inhomogeneous DAs. The ratio $c_{121}/c_{120}$ is calculated, and errors  are given. Note that the error in the ratios is typically less than $1/50$th the error in the coefficients. }
\begin{center}
 \scalebox{0.9}{
\begin{tabular}{|ccccccc|}
\hline \\
Approximant & $c_{120} $ & Error & $c_{121}$ & Error & Ratio & Error \\
\hline
Exact & $3.85433525\ldots e69$ & 0 & $1.44743231\ldots e70$ & 0 & $3.75533579\ldots $&0 \\
$[4,3,5,4,1]$ & $ 3.85409328e69$ & $6.4e-5$ & $1.44733975e70$& $6.3e-5$ & $3.75533140$ & $1.2e-6$ \\
$[4,4,4,4,1]$ & $ 3.85432477e69$ & $2.7e-6$ & $1.44742835e70$& $2.7e-5$ & $3.7553357$ & $1.9e-8$ \\
$[4,5,3,4,1]$ & $ 3.85402107e69$ & $8.3e-5$ & $1.44731207e70$& $8.2e-5$ & $3.7553301$ & $7.6e-7$ \\
$[4,3,4,4,2]$ & $ 3.85419555e69$ & $3.6e-5$ & $1.4473705e70$& $3.6e-5$ & $3.7553341$ & $4.5e-7$ \\
$[4,4,3,4,2]$ & $ 3.85802613e69$ & $9.7e-4$ & $1.44885029e70$& $9.6e-4$ & $3.7554186$ & $2.2e-5$ \\
$[4,3,3,4,3]$ & $ 3.85328878e69$ & $2.8e-4$ & $1.44703211e70$& $2.7e-4$ & $3.7553171$ & $5.0e-6$ \\
\hline

\end{tabular}
}
\end{center}
\label{isingt20a}
\end{table}%

\begin{table}[!ht]
\caption{Prediction of $c_{240} $ and $c_{241}$ from triangular Ising susceptibility series to order $z^{40}$ by third order inhomogeneous DAs. The ratio $c_{241}/c_{240}$ is calculated, and errors  are given. Note that the error in the ratios is typically less than $1/50$th the error in the coefficients.}
\begin{center}
 \scalebox{0.8}{
\begin{tabular}{|ccccccc|}
\hline \\
Approximant & $c_{240} $ & Error & $c_{241}$ & Error & Ratio & Error \\
\hline
Exact & $2.788143197746\ldots e138$ & 0 & $1.04379811183\ldots e139$ & 0 & $3.743703381791\ldots $&0 \\
$[9,8,10,9,1]$ & $2.78814318268e138$ & $5.4e-9$ & $1.04379810612e139$ & $5 .5e-9$ & $3.74370338154$ &  $6.7e-11$ \\
$[9,9,9,9,1]$  &  2.78814319431e138 &  1.2e-9 &  1.04379811053e139 &  1.3e-9 &  3.74370338173 &  1.6e-11\\
$[9,10,8,9,1]$& 2.78814319386e138 &  1.4e-9 &  1.04379811036e139 &  1.4e-9 &  3.74370338172 &  1.8e-11 \\
$[9,8,9,9,2]$  &  2.78814319672e138 &  3.7e-10 &  1.04379811144e139 &  3.7e-10 &  3.74370338177 &  5.2e-12 \\
$[9,9,8,9,2]$  &  2.78814319369e138 &  1.5e-9 &  1.04379811029e139 &  1.5e-9 &  3.74370338172 &  1.9e-11 \\
$[8,9,9,8,3]$&2.78814319962e138 &  6.7e-10 &  1.04379811254e139 &  6.8e-10 &  3.74370338182 &  7.5e-12 \\
$[9,8,8,9,3]$  &  2.78814320043e138 &  9.6e-10 &  1.04379811285e139 &  9.7e-10 &  3.74370338183 &  1.1e-11 \\
$[9,8,8,9,3]$  &  2.78814319036e138 &  2.6e-9 &  1.04379810903e139 &  2.7e-9 &  3.74370338167 &  3.3e-11 \\
$[8,8,9,8,4]$ &  2.78814320724e138 &  3.4e-9 &  1.04379811543e139 &  3.4e-9 &  3.74370338195 &  4.1e-11\\
$[8,9,8,8,4]$ &  2.78814320682e138 &  3.3e-9 &  1.04379811527e139 &  3.3e-9 &  3.74370338194 &  4.0e-11\\
$[8,8,8,8,5]$ &  2.78814320688e138 &  3.3e-9 &  1.04379811529e139 &  3.3e-9 &  3.74370338194 &  4.0e-11\\
$[8,9,7,8,5]$  &  2.78814320967e138 &  4.3e-9 &  1.04379811635e139 &  4.3e-9 &  3.74370338199 &  5.4e-11\\
$[8,7,8,8,6]$&  2.78814319603e138 &  6.2e-10 &  1.04379811118e139 &  6.3e-10 &  3.74370338176 &  8.1e-12\\
$[8,8,7,8,6]$  &  2.78814319910e138 &  4.8e-10 &  1.04379811234e139 &  4.9e-10 &  3.74370338181 &  5.5e-12\\
$[7,8,8,7,7]$ &  2.78814319036e138 &  2.6e-9 &  1.04379810903e139 &  2.7e-9 &  3.74370338167 &  3.3e-11\\
$[8,7,7,8,7]$ &  2.78814319599e138 &  6.3e-10 &  1.04379811117e139 &  6.4e-10 &  3.74370338176 &  8.5e-12\\
$[7,7,8,7,8]$  &  2.78814318147e138 &  5.8e-9 &  1.04379810567e139 &  5.9e-9 &  3.74370338155 &  6.5e-11\\
$[7,8,7,7,8]$ &  2.78814318814e138 &  3.4e-9 &  1.04379810819e139 &  3.5e-9 &  3.74370338163 &  4.2e-11\\
\hline

\end{tabular}
}
\end{center}
\label{isingt40}
\end{table}%

We now consider the extent to which this series extension improves the quality of the ratio analysis. From the above data we show in figure 2 the estimators $\mu_n$ of the growth constant $\mu,$ as defined by (\ref{muest}), obtained from the given 20 coefficients of the triangular Ising susceptibility series. The exact growth constant is $2+\sqrt{3}=3.73205\cdots$ . Some curvature is evident, and linearly extrapolating the last points of the plot, one might estimate $\mu \approx 3.732.$ In figure 3 the corresponding estimators of  $\mu_n$ obtained from the next 100 approximate coefficients of the series are shown. Note the scale of the ordinate is now reduced by an order of magnitude.  Linearly extrapolating the last points of the plot, one might estimate $\mu \approx 3.73205,$ an improvement in both accuracy and precision by two orders-of-magnitude.

In  figure 4 the estimators $\gamma_n$ of the exponent $\gamma,$ as defined by (\ref{gamest}), (recall the exact value of $z_c$ is assumed in this case), obtained from the given 20 coefficients of the triangular Ising susceptibility series are shown. The exact exponent  is $7/4.$  Again some curvature is evident, and linearly extrapolating the last points of the plot, one might estimate $\gamma \approx 1.750.$ In figure 5 the corresponding estimators of the exponent  obtained from the next 100 approximate coefficients of the series are shown. Note the scale of the ordinate is now reduced threefold, and the plot is much more linear.  Linearly extrapolating the last points of the plot, one might estimate $\gamma \approx 1.7500,$ an order-of-magnitude improvement in both accuracy and precision.

This series however is a canonical example of the sort of series that is ideally suited to analysis by differential approximants. Using the same 20 terms, 2nd order DAs allow one to estimate the radius of convergence as $z_c = 0.267951 \pm 0.000005,$ and the exponent as $\gamma=1.7502 \pm 0.0006.$ With 40 terms the DAs allow one to estimate the radius of convergence and exponent with errors confined to the 11th and 8th significant digits respectively. Our point in this example is not to argue that the ratio method is, or can be made, superior to the DA method for the estimate of the critical point and critical exponent, but rather to show that it can significantly improve the performance of the ratio method, and also predict a surprising number of coefficients. 

Furthermore, it does provide a substantial improvement in the estimate of the critical amplitude. The amplitude is $C/\Gamma(7/4) = 0.9216808677\cdots,$ \cite{C10}. Using the 40 term susceptibility series, and extrapolating the last two estimators $C_n$ as given by (\ref{eq:amp}) against $1/n,$ one obtains the estimate $C/\Gamma(7/4) = 0.921728\cdots,$ which is in error by $0.000043.$ Now, using th DAs to extend the series by 200 terms as discussed, the extrapolated amplitude from the last few (extrapolated) coefficients is $C/\Gamma(7/4) = 0.9216822\cdots,$ so the error is reduced to $0.0000013,$ giving an improvement in accuracy by a factor of more than 30.

\setlength{\unitlength}{1.5cm}
\begin{picture}(2,4) \label{fig:isingt1}
\put(2,-1.8){\includegraphics[height=8cm]{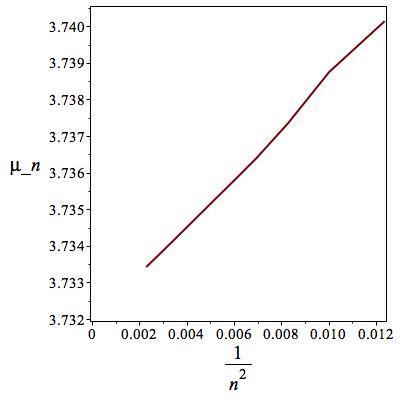}}
\put(0,-2.4){Fig. 2: Plot of estimators of $\mu$ against $1/n^2$ from the first 20 exact coefficients of the }
\put(0,-2.75){triangular lattice Ising susceptibility series. The exact value is $\mu=2+\sqrt{3}=3.73205\ldots.$}
\end{picture}

 \begin{table}[H]
\caption{Prediction of $c_{240} $ and $c_{241}$ from triangular Ising susceptibility series to order $z^{40}$ by fourth order inhomogeneous DAs. The ratio $c_{241}/c_{240}$ is calculated, and errors  are given. Note that the error in the ratios is typically less than $1/2$ the error in the coefficients.}
\begin{center}
\scalebox{0.85}{
\begin{tabular}{|ccccccc|}
\hline \\
Approximant & $c_{240} $ & Error & $c_{241}$ & Error & Ratio & Error \\
\hline
Exact & $2.788143197746\ldots e138$ & 0 & $1.04379811183\ldots e139$ & 0 & $3.743703381791\ldots $&0 \\
$[7,6,8,7,7,1]$&  2.78814340767e138 &  7.5e-8 &  1.04379816691e139 &  5.3e-8 &  3.74370329746 &  2.3e-8 \\
$[7,7,7,7,7,1]$ &  2.78814340772e138 &  7.5e-8 &  1.04379816684e139 &  5.3e-8 &  3.74370329715 &  2.3e-8\\
$[7,8,6,7,7,1]$ &  2.78814339248e138 &  7.0e-8 &  1.04379815848e139 &  4.5e-8 &  3.74370328762 &  2.5e-8\\
 $[7,6,7,7,7,2]$&  2.78814344604e138 &  8.9e-8 &  1.04379818102e139 &  6.6e-8 &  3.74370329655 &  2.3e-8\\
$[7,7,6,7,7,2]$ &  2.78814440196e138 &  4.3e-7 &  1.04379853946e139 &  4.1e-7 &  3.74370329859 &  2.2e-8\\
$[7,6,6,7,7,3]$ &  2.78815532086e138 &  4.3e-6 &  1.04380263067e139 &  4.3e-6 &  3.74370331114 &  1.9e-8\\
$[6,7,7,6,6,4]$  &  2.78814327640e138 &  2.8e-8 &  1.04379811767e139 &  5.6e-9 &  3.74370329713 &  2.3e-8\\
$[6,7,6,6,6,5]$  &  2.78814336359e138 &  5.9e-8 &  1.04379815082e139 &  3.7e-8 &  3.74370329895 &  2.2e-8\\
$[6,6,6,6,6,6]$ &  2.78814340974e138 &  7.6e-8 &  1.04379816817e139 &  5.4e-8 &  3.74370329920 &  2.2e-8\\
$[6,7,5,6,6,6]$  &  2.78814347427e138 &  9.9e-8 &  1.04379819163e139 &  7.6e-8 &  3.74370329669 &  2.3e-8\\
$[6,5,6,6,6,7]$   &  2.78814337576e138 &  6.4e-8 &  1.04379815552e139 &  4.2e-8 &  3.74370329945 &  2.2e-8\\
$[6,6,5,6,6,7]$   &  2.78814338706e138 &  6.8e-8 &  1.04379815977e139 &  4.6e-8 &  3.74370329952 &  2.2e-8\\
$[6,5,5,6,6,8]$ &  2.78814338299e138 &  6.6e-8 &  1.04379815760e139 &  4.4e-8 &  3.74370329721 &  2.3e-8 \\

\hline

\end{tabular}
}
\end{center}
\label{isingt40a}
\end{table}%

\setlength{\unitlength}{1.5cm}
\begin{picture}(2,4) \label{fig:isingt2}
\put(2,-1.8){\includegraphics[height=8cm]{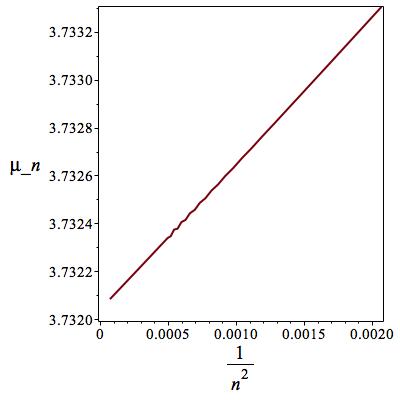}}
\put(0,-2.4){Fig. 3: Plot of  estimators of $\mu$ against $1/n^2$ from the approximate coefficients $c_{21}$ to $c_{120}$  }
\put(0,-2.75){of the triangular lattice Ising susceptibility series. The exact value is $\mu=2+\sqrt{3}=3.73205\ldots.$}
\end{picture}

\setlength{\unitlength}{1.5cm}
\begin{picture}(2,4) \label{fig:isingt3}
\put(2,-1.8){\includegraphics[height=8cm]{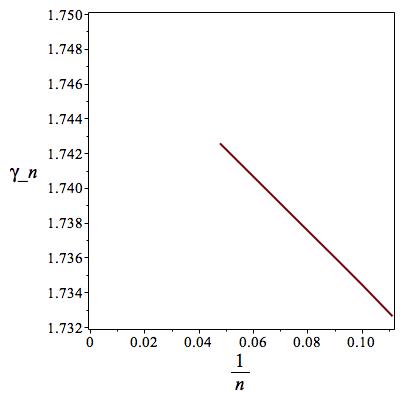}}
\put(0,-2.4){Fig. 4: Plot of estimators of the exponent $\gamma$ against $1/n$ from the first 20 coefficients of }
\put(0,-2.7){the triangular lattice Ising susceptibility series. The exact value is $\gamma=1.75.$}
\end{picture}

\setlength{\unitlength}{1.5cm}
\begin{picture}(2,4) \label{fig:isingt4}
\put(2,-4.8){\includegraphics[height=8cm]{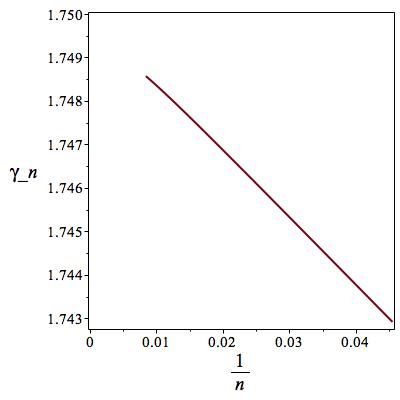}}
\put(0,-5.25){Fig. 5: Plot of estimators of the exponent $\gamma$ against $1/n$ from the approximate coefficients   }
\put(0,-5.55){$c_{21}$ to $c_{120}$  of the triangular lattice Ising susceptibility series. The exact value is $\gamma=1.75.$}
\end{picture}

 \begin{table}[H]
\caption{Prediction of $c_{120} $ and $c_{121}$ from square Ising susceptibility series to order $z^{20}$ by second and third order inhomogeneous DAs. The ratio $c_{121}/c_{120}$ is calculated, and errors  are given. Note that the error in the ratios is typically comparable to the error in the coefficients.}
\begin{center}
 \scalebox{0.85}{
\begin{tabular}{|ccccccc|}
\hline \\
Approximant & $c_{120} $ & Error & $c_{121}$ & Error & Ratio & Error \\
\hline
Exact & $3.699344911656\ldots e37$ & 0 & $7.444983988658\ldots e37$ & 0 & $2.01251415222\ldots $&0 \\
$[5,6,6,1]$  &  3.69911318185e37 &  6.3e-5 &  7.44461366194e37 &  5.0e-5 &  2.01254010201 &  1.3e-5 \\
$[6,6,5,1]$  &  3.71823472589e37 &  5.1e-3 &  7.47065449207e37 &  3.4e-3 &  2.00919389725 &  1.6e-3\\
$[5,5,6,2]$ &   3.69553924958e37 &  1.0e-3 &  7.43725054250e37 &  1.0e-3 &  2.01249408722 &  1.0e-5  \\
$[6,5,5,2]$ &   3.69256409470e37 &  1.8e-3 &  7.41477918329e37 &  4.1e-3 &  2.00802993774 &  2.2e-3\\
$[5,4,6,3]$ &   3.70862788066e37 &  2.5e-3 &  7.46384656529e37 &  2.5e-3 &  2.01256275177 &  2.4e-5 \\
$[5,5,5,3]$  &  3.73883726557e37 &  1.1e-2 &  7.47810929156e37 &  4.4e-3 &  2.00011634827 &  6.2e-3 \\
$[5,6,4,3]$ &  3.72918712388e37 &  8.1e-3 &  7.47514696158e37 &  4.1e-3 &  2.00449776649 &  4.0e-3 \\
$[5,5,4,4]$  &  3.73888615283e37 &  1.1e-2 &  7.41799657976e37 &  3.6e-3 &  1.98401236534 &  1.4e-2 \\
$[4,5,4,5]$  &  3.60362014492e37 &  2.6e-2 &  7.27064854332e37 &  2.3e-2 &  2.01759576797 &  2.5e-3 \\
$[5,4,4,5]$&  4.01836223688e37 &  8.6e-2 &  6.95314516058e37 &  6.6e-2 &  1.73034310341 &  1.4e-1 \\
$[3,3,4,8]$  &  3.52503862309e37 &  4.7e-2 &  7.09241886841e37 &  4.7e-2 &  2.01201176643 &  2.5e-4 \\
$[4,3,5,4,1]$  &  3.72255494919e37 &  6.3e-3 &  7.46715119902e37 &  3.0e-3 &  2.00592112541 &  3.3e-3\\
$[4,4,4,4,1]$  &  3.70120934213e37 &  5.0e-4 &  7.44889682792e37 &  5.3e-4 &  2.01255750656 &  2.2e-5 \\
\hline
\end{tabular}
}
\end{center}
\label{isings20}
\end{table}

\begin{table}[H]
\caption{Prediction of $c_{140} $ and $c_{141}$ from square Ising susceptibility series to order $z^{40}$ by third order inhomogeneous DAs. The ratio $c_{141}/c_{140}$ is calculated, and errors  are given. Note that the error in the ratios is typically comparable to the error in the coefficients.}
\begin{center}
 \scalebox{0.85}{
\begin{tabular}{|ccccccc|}
\hline \\
Approximant & $c_{140} $ & Error & $c_{141}$ & Error & Ratio & Error \\
\hline
Exact & $4.3564015730407\ldots e45$ & 0 & $8.7595244423367\ldots e45$ & 0 & $2.0107247450612\ldots $&0 \\
$[9,8,10,9,1]$  &  4.35639884330e43 &  6.3e-7 &  8.75951476396e43 &  1.1e-6 &  2.01072382927 &  4.7e-7\\
$[9,9,9,9,1]$   &  4.35640131225e43 &  6.0e-8 &  8.75952119831e43 &  3.7e-7 &  2.01072406769 &  3.6e-7\\
$[9,10,8,9,1]$ &  4.35640554206e43 &  9.1e-7 &  8.75951506541e43 &  1.1e-6 &  2.01072072983 &  2.0e-6 \\
$[9,8,9,9,2]$  &  4.35640334479e43 &  4.1e-7 &  8.75951049168e43 &  1.6e-6 &  2.01072072983 &  2.0e-6 \\
$[9,9,8,9,2]$  &  4.35640350448e43 &  4.4e-7 &  8.75950944587e43 &  1.7e-6 &  2.01072049141 &  2.1e-6 \\
$[8,9,9,8,3]$ &  4.35641341130e43 &  2.7e-6 &  8.75952195353e43 &  2.8e-7 &  2.01071882248 &  3.0e-6 \\
$[9,8,8,9,3]$   &  4.35640428083e43 &  6.2e-7 &  8.75950768594e43 &  1.9e-6 &  2.01071953773 &  2.6e-6 \\
$[8,9,8,8,4]$   &  4.35639824915e43 &  7.6e-7 &  8.75946120016e43 &  7.2e-6 &  2.01071166992 &  6.5e-6\\
$[8,8,8,8,5]$  &  4.35639714447e43 &  1.0e-6 &  8.75952509717e43 &  7.5e-8 &  2.01072692871 &  1.1e-6\\
$[8,9,7,8,5]$  &  4.35640191254e43 &  7.8e-8 &  8.75951072676e43 &  1.6e-6 &  2.01072144508 &  1.7e-6\\
$[8,7,8,8,6]$ &  4.35635226526e43 &  1.1e-5 &  8.75943511456e43 &  1.0e-5 &  2.01072692871 &  1.1e-6\\
$[8,8,7,8,6]$   &  4.35640136885e43 &  4.7e-8 &  8.75950037613e43 &  2.7e-6 &  2.01071929932 &  2.7e-6\\
$[8,7,7,8,7]$ &  4.35640325285e43 &  3.9e-7 &  8.75950427151e43 &  2.3e-6 &  2.01071929932 &  2.7e-6\\
$[7,7,8,7,8]$   &  4.35640143172e43 &  3.2e-8 &  8.75957205963e43 &  5.4e-6 &  2.01073575020 &  5.5e-6\\
\hline

\end{tabular}
}
\end{center}
\label{isings40}
\end{table}%

\subsection{Three-dimensional convex polygon series}
In the above examples, the underlying susceptibility function is widely believed to be non-D-finite, as evidenced by a natural boundary in the complex plane. While there is overwhelming numerical evidence for this \cite{C10}, it has not been proved. Another example which is provably non-D-finite is the generating function for three-dimensional convex polygons, obtained by Bousquet-M\'elou and Guttmann \cite{BMG96}, who gave the closed form expression for the generating function. We have expanded this, and used the first 38 terms in the expansion to predict the next 100 terms from DAs. In table \ref{convex3} we give the predicted coefficients $c_{136}$ and $c_{137}$ which are seen to be predicted to an accuracy of a few parts in $10^8$ or $10^9,$ while the {\em ratios} are typically predicted with a precision at least 10 times better than that of the coefficients themselves.
 \begin{table}[H]
\caption{Prediction of $c_{137} $ and $c_{138}$ from the series for three-dimensional convex polygons to order $z^{38}$ by third order inhomogeneous DAs. The ratio $c_{138}/c_{137}$ is calculated, and errors  are given. Note that the error in the ratios is typically less than $1/10$th the error in the coefficients.}
\begin{center}
\scalebox{0.82}{
\begin{tabular}{|ccccccc|}
\hline \\
Approximant & $c_{136} $ & Error & $c_{137}$ & Error & Ratio & Error \\
\hline
Exact & $1.32618751844344\ldots e134$ & 0 & $1.2110917705117\ldots e135$ & 0 & $9.132130665301268\ldots $&0 \\
$[8,8,9,8,1]$  &  1.32618748924e134 &  2.2e-8 &  1.21109174109e135 &  2.4e-8 &  9.13213064453 &  2.3e-9\\
$[8,9,8,8,1]$  &     1.32618748869e134 &  2.2e-8 &  1.21109174052e135 &  2.5e-8 &  9.13213064407 &  2.3e-9\\
 $[8,7,9,8,2]$ &  1.32618567633e134 &  1.4e-6 &  1.21108975408e135 &  1.7e-6 &  9.13212814540 &  2.8e-7\\
$[8,8,8,8,2]$ &  1.32618567253e134 &  1.4e-6 &  1.21108974983e135 &  1.7e-6 &  9.13212813951 &  2.8e-7\\
$[8,9,7,8,2]$ &  1.32618343126e134 &  3.1e-6 &  1.21108721811e135 &  3.8e-6 &  9.13212448268 &  6.8e-7\\
$[8,7,8,8,3]$   &  1.32618568738e134 &  1.4e-6 &  1.21108976643e135 &  1.7e-6 &  9.13212816236 &  2.7e-7\\
$[8,8,7,8,3]$    &  1.32618086853e134 &  5.0e-6 &  1.21108430134e135 &  6.2e-6 &  9.13212013594 &  1.2e-6\\
$[7,8,8,7,4]$  &  1.32618819358e134 &  5.1e-7 &  1.21109251442e135 &  6.1e-7 &  9.13213162567 &  1.1e-7\\
$[7,7,8,7,5]$   &  1.32618752590e134 &  5.6e-9 &  1.21109177783e135 &  6.0e-9 &  9.13213066913 &  4.2e-10\\
$[7,8,7,7,5]$   &  1.32618752596e134 &  5.7e-9 &  1.21109177789e135 &  6.1e-9 &  9.13213066916 &  4.2e-10\\
$[7,6,8,7,6]$    &  1.32618752289e134 &  3.4e-9 &  1.21109177489e135 &  3.6e-9 &  9.13213066772 &  2.7e-10\\
$[7,7,7,7,6]$  &  1.32618752234e134 &  2.9e-9 &  1.21109177436e135 &  3.2e-9 &  9.13213066745 &  2.4e-10\\
$[7,8,6,7,6]$  &  1.32618752276e134 &  3.3e-9 &  1.21109177477e135 &  3.5e-9 &  9.13213066766 &  2.6e-10 \\
$[7,6,7,7,7]$   &  1.32618752259e134 &  3.1e-9 &  1.21109177460e135 &  3.4e-9 &  9.13213066757 &  2.5e-10\\
$[7,7,6,7,7]$   &  1.32618752281e134 &  3.3e-9 &  1.21109177481e135 &  3.5e-9 &  9.13213066767 &  2.6e-10\\
$[6,7,7,6,8]$   &  1.32618752235e134 &  2.9e-9 &  1.21109177436e135 &  3.2e-9 &  9.13213066744 &  2.3e-10\\
$[7,6,6,7,8]$ &  1.32618752259e134 &  3.1e-9 &  1.21109177460e135 &  3.4e-9 &  9.13213066758 &  2.5e-10\\
\hline

\end{tabular}
}
\end{center}
\label{convex3}
\end{table}%

As with the triangular susceptibility data, we investigated the improvement that this series extension affords for a ratio analysis of this series. We show in figure 6 the estimators $\mu_n$ of the growth constant $\mu,$ as defined by (\ref{muest}), obtained from the first 38 exact coefficients of the series. The exact growth constant is exactly $9.0.$  A lot of curvature is evident, primarily because the  asymptotic form of the coefficients is  $const. \,\, 9^n \,\,n^2 \left (1 + \frac{c_1}{n} + \frac{c_2}{n\log n} +  \frac{c_2}{n\log^2 n} + \ldots \right ).$ Linearly extrapolating the last points of the plot, one might estimate $\mu =9.00 \pm 0.02.$ In figure 7 the corresponding estimators $\mu_n$ of the growth constant $\mu$ obtained from the next 100 approximate coefficients of the series are shown. Note the scale of the ordinate is now reduced by more than an order of magnitude.  Linearly extrapolating the last points of the plot, one might estimate $\mu =9.000 \pm 0.002,$ an improvement of an order-of-magnitude in precision.

In  figure 8 the estimators $\gamma_n$ of the exponent $\gamma,$ as defined by (\ref{gamest}), obtained from the first 38 exact coefficients of the series are shown. The exact exponent  is $3.0.$  Again considerable curvature is evident, and linearly extrapolating the last points of the plot, one might estimate $\gamma = 3.0 \pm 0.1.$ In figure 9 the corresponding estimators $\gamma_n$  obtained from the next 100 approximate coefficients of the series are shown. Note the scale of the ordinate is now reduced by nearly an order of magnitude, and the plot is much more linear.  Linearly extrapolating the last points of the plot, one might estimate $\gamma =3.00 \pm 0.01,$ an order-of-magnitude improvement in precision.

Again, a higher level of precision is afforded by a DA analysis, though the results need rather more careful analysis than in the previos example. The estimates of the critical point, using 2nd order DAs, suggest $z_c=0.11111101 \pm 0.00000001,$ while the exponent estimate is $\gamma=-2.9992 \pm 0.0001.$ The error bars are seen to be an order of magnitude too small to include the exact results. This is due to the presence of sub-dominant logarithmic terms in the asymptotics (see above), and these are hinted at in the DAs by the presence of nearby singularities at $z \approx 0.112\cdots,$ and $z \approx 0.117 \cdots.$ As in the previous example, the extrapolated series coefficients substantially reduce the error in the estimate of the dominant amplitude compared to that obtained from the known series coefficients.

\vspace{-6mm}
\setlength{\unitlength}{1.5cm}
\begin{picture}(2,4) \label{fig:convex1}
\put(2,-1.8){\includegraphics[height=8cm]{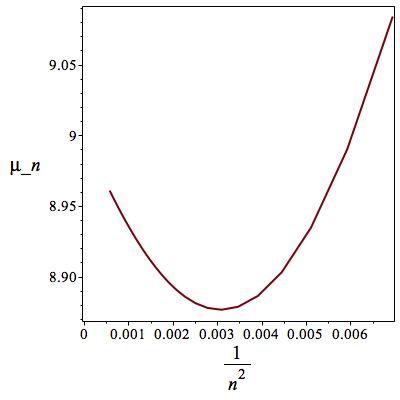}}
\put(0,-2.1){Fig. 6: Plot of  estimators of $\mu$ against $1/n^2$ from the first 38 exact coefficients }
\put(0,-2.4){of the three-dimensional convex polygon series. The exact value of $\mu$ is $9.$}
\end{picture}

\setlength{\unitlength}{1.5cm}
\begin{picture}(2,4) \label{fig:convex2}
\put(2,-1.8){\includegraphics[height=8cm]{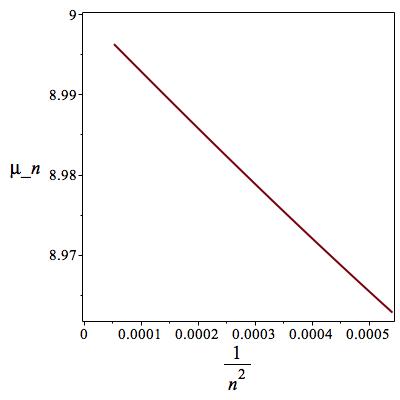}}
\put(0,-2.2){Fig. 7: Plot of estimators of $\mu$ against $1/n^2$ from the approximate coefficients $c_{39}$   }
\put(0,-2.5){to $c_{138}$ of the three-dimensional convex polygon series. The exact value of $\mu$ is $9.$}
\end{picture}

\setlength{\unitlength}{1.5cm}
\begin{picture}(2,4) \label{fig:convex33}
\put(2,-4.8){\includegraphics[height=8cm]{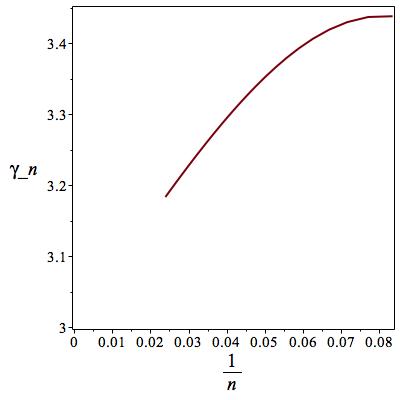}}
\put(0,-5.25){Fig. 8: Plot of estimators of the exponent $\gamma$ against $1/n$ from the first 38 exact }
\put(0,-5.55){coefficients of the three-dimensional convex polygon series. The exact value is $\gamma=3.$}
\end{picture}

\newpage

\setlength{\unitlength}{1.5cm}
\begin{picture}(2,4) \label{fig:convex4}
\put(2,-1.){\includegraphics[height=8cm]{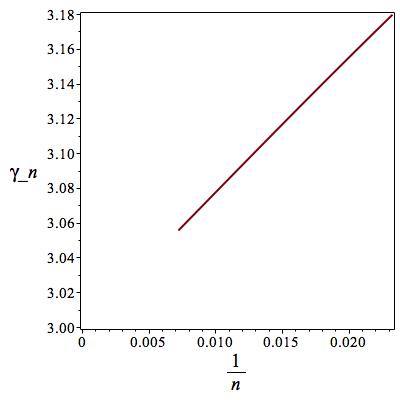}}
\put(0,-1.45){Fig. 9: Estimators of the exponent $\gamma$ against $1/n$ from the approximate coefficients $c_{39}$   }
\put(0,-1.75){ to $c_{138}$ of the three-dimensional convex polygon series. The exact value is $\gamma=3.$}
\end{picture}

\vspace{3cm}

\subsection{Functions with non-algebraic singularities}

For all the above problems, the asymptotic form of the coefficients is believed to be that of an algebraic singularity. That is to say, the OGF is believed to behave as $$f(z) \sim C \cdot (1-\mu \cdot z)^{-\gamma}.$$ Consequently, the $n^{th}$ coefficient in the OGF has asymptotic behaviour $const \cdot \mu^n \cdot n^{\gamma-1}.$ Note that this form is ideally suited to analysis by the DA method, which, in such cases, usually gives excellent estimates of $\mu,$ and good estimates of $\gamma,$ as we have seen.

There is quite a large class of problems for which the asymptotic form of the coefficients includes a stretched exponential term, so that the coefficients behave as $$const \cdot \mu^n\cdot \mu_1^{n^\sigma} \cdot n^{\gamma-1},$$ and $\mu_1 < 1.$ For such functions the method of differential approximates does a very poor job of estimating the critical point and critical exponent. The estimate of $\mu$ is typically only accurate to 2, 3 or 4 digits, rather than 10-20 digits in the case of an algebraic singularity. Moreover, the exponent estimate is completely unreliable, varying in both sign and magnitude between DAs. Examples of such behaviour are given in \cite{G14}. This is hardly surprising, as functions satisfying the assumed linear, inhomogeneous ODE cannot have asymptotics with this stretched exponential behaviour. 

Given how badly DAs approximate the singular behaviour,  the idea of using the DAs to extend the known series in this case may, at first sight, seem like a complete waste of time. 
To check this, we took the series for pushed Dyck paths. These are Dyck paths with an additional variable $y$ associated with the height of the highest vertex. One then has a two-variable generating function, with one variable $z$ conjugate to the length of the path and the other variable $y$ conjugate to the maximum height. If $y<1$, Dyck paths with small maximal height are favoured, and if $y > 1$ tall Dyck paths are favoured. We refer to these situations as {\em pushed} and {\em pulled} respectively. 

We took for our example Dyck paths with pushing fugacity $y=0.5$, for which we have 2000 coefficients \cite{G14}. The series coefficients have the asymptotic form $const \cdot \mu^n\cdot \mu_1^{n^{1/3}} \cdot n^{\gamma-1},$ with $\mu=4,$ $\gamma=1/6$ and $\mu_1 < 1.$ We took the first 40 terms and predicted the next 50 terms. These agreed to 24 significant digits (machine precision) for the first predicted term i.e. $c_{41},$ down to 8 significant digits for the 50th predicted term, $c_{90}.$ More significantly, the predictions are correct, with errors in every case being confined to a few parts in the last quoted digit. The predicted and actual coefficients are shown in table \ref{tab:booktabs} below. More precisely, the left hand column gives the predicted coefficients of $z^{41}$ to $z^{90}$ for pushed Dyck paths with $y=0.5,$ while the right hand column gives the exact (up to rounding) values of the coefficients.

Finally, having shown we can predict 50 further terms from a 40 term series, we try and push this a little further. We take a very slightly shorter series of 38 terms for the same problem, and  predict 100 further terms. The results are shown in table \ref{dyck100}, where it can be seen that the predicted coefficients are obtained with a precision of a few parts in $10^4$ and the ratios are predicted with a precision of a few parts in $10^5.$ This is accurate enough for a ratio analysis.

So it appears that, while the DA method definitely does a poor job in predicting the critical point and exponent, it does a surprisingly good job of approximating the coefficients. This is a very significant observation, as many problems with coefficients displaying this stretched exponential behaviour present considerable challenges in computing the coefficients (for example, 1324 pattern-avoiding permutations \cite{CG14}), and furthermore the ratio method is very useful in studying the asymptotics of such problems.

\begin{table}
 \caption{Predicted and actual coefficients of $z^{41}$ to $z^{90}$ for pushed Dyck paths with $y=0.5.$}
\begin{center}
 \scalebox{0.9}{
   \begin{tabular}{@{} lc @{}} 
      \hline
      Predicted    & Actual coefficient\\
\hline
      1.06242311170274760152467e+20&1.062423111702747601524648e20\\
3.825403274070053801475e+20 &   3.8254032740700538014727e+20 \\
1.3797869385954645590197e+21 &   1.3797869385954645590109e+21 \\
4.9850845560541723609e+21 &   4.9850845560541723606784e+21 \\
1.803977425231757837e+22 &   1.8039774252317578369439e+22 \\
6.53825324627817524e+22 &   6.5382532462781752272489e+22 \\
2.37323063928703901e+23 &   2.3732306392870389970783e+23 \\
8.6266584907208658e+23 &   8.6266584907208655572576e+23 \\
3.1401317884469481e+24 &   3.1401317884469478637918e+24 \\
1.144550869280430e+25 &   1.1445508692804297946738e+25 \\
4.177194711962482e+25 &   4.1771947119624793925414e+25 \\
1.526434580362285e+26 &   1.5264345803622824558865e+26 \\
5.58467543172355e+26 &   5.5846754317235197201892e+26 \\
2.04563071867204e+27 &   2.0456307186720238819845e+27 \\
7.5015361625622e+27 &   7.5015361625619753870851e+27 \\
2.7539226184982e+28 &   2.7539226184980997986724e+28 \\
1.0120859980001e+29 &   1.0120859980000582564781e+29 \\
3.7233471756384e+29 &   3.7233471756375094820673e+29 \\
1.3711563779688e+30 &   1.3711563779681653078966e+30 \\
5.054349616683e+30 &   5.0543496166785380877423e+30 \\
1.864903911635e+31 &   1.8649039116326064384071e+31 \\
6.88730165917e+31 &   6.8873016591526046477679e+31 \\
2.54584689945e+32 &   2.5458468994390695789213e+32 \\
9.41879887642e+32 &   9.4187988763396211635697e+32 \\
3.48761706915e+33 &   3.4876170691095912329179e+33 \\
1.29247596738e+34 &   1.2924759673592608945370e+34 \\
4.79366122770e+34 &   4.7936612275293845584592e+34 \\
1.77932188150e+35 &   1.7793218814039961065336e+35 \\
6.6096049891e+35 &   6.6096049885464199499817e+35 \\
2.4570959331e+36 &   2.4570959328721307481647e+36 \\
9.140847682e+36 &   9.1408476807503133947131e+36 \\
3.402992634e+37 &   3.4029926330368814413056e+37 \\
1.267764048e+38 &   1.2677640476976414057377e+38 \\
4.726195864e+38 &   4.7261958610467492931807e+38 \\
1.763088413e+39 &   1.7630884110366671003139e+39 \\
6.581411255e+39 &   6.5814112476992118944904e+39 \\
2.458330294e+40 &   2.4583302907361180747460e+40 \\
9.18822503e+40 &   9.1882250096878304522944e+40 \\
3.43627038e+41 &   3.4362703769728698807864e+41 \\
1.285883672e+42 &   1.2858836679511071317055e+42 \\
4.81469965e+42 &   4.8146996269989242045569e+42 \\
1.803784229e+43 &   1.8037842171993073337432e+43 \\
6.76149470e+43 &   6.7614946433744471487360e+43 \\
2.53593793e+44 &   2.5359379066324063018430e+44 \\
9.5162865e+44 &   9.5162863745776859619999e+44 \\
3.57293182e+45 &   3.5729317525743742288111e+45 \\
1.34216458e+46 &   1.3421645526852449078560e+46 \\
5.04436242e+46 &   5.0443622768736697735393e+46 \\
1.89680208e+47 &   1.8968020147767692514266e+47 \\
7.1359020e+47 &   7.1359017074535129004836e+47 \\

      \hline
   \end{tabular}
   \label{tab:booktabs}
}
\end{center}
\end{table}

 \begin{table}[H]
\caption{Prediction of $c_{136} $ and $c_{137}$ from pushed Dyck paths series, with force fugacity $y=0.5,$ to order $z^{37}$ by third order inhomogeneous DAs. The ratio $c_{137}/c_{136}$ is calculated, and errors  are given. Note that the error in the ratio is typically less than $1/10$th the error in the coefficients.}
\begin{center}
\scalebox{0.9}{
\begin{tabular}{|ccccccc|}
\hline \\
Approximant & $c_{136} $ & Error & $c_{137}$ & Error & Ratio & Error \\
\hline
Exact & $8.2649688767175\ldots e73$ & 0 & $3.1578959225337\ldots e75$ & 0 & $3.8208201018\ldots $&0 \\
$[8,7,9,8,1]$&  8.37439989184e73 &  1.3e-2 &  3.20534771370e74 &  1.5e-2 &  3.82755517960 &  1.8e-3 \\
$[8,8,8,8,1]$  &  8.26193395085e73 &  3.7e-4 &  3.15666749323e74 &  3.9e-4 &  3.82073688507 &  2.2e-5\\
$[8,9,7,8,1]$ &  8.46996643612e73 &  2.5e-2 &  3.24771410878e74 &  2.8e-2 &  3.83438849449 &  3.6e-3\\
 $[8,7,8,8,2]$&  8.26821563575e73 &  3.9e-4 &  3.15923396586e74 &  4.2e-4 &  3.82093811035 &  3.1e-5\\
$[8,8,7,8,2]$  &  8.26869458177e73 &  4.5e-4 &  3.15943314267e74 &  4.9e-4 &  3.82095766068 &  3.6e-5\\
$[7,8,8,7,3]$  &  8.26717229783e73 &  2.7e-4 &  3.15880163688e74 &  2.9e-4 &  3.82089734077 &  2.0e-5\\
$[7,7,8,7,4]$   &  8.26728344922e73 &  2.8e-4 &  3.15884774467e74 &  3.0e-4 &  3.82090163231 &  2.1e-5\\
$[7,8,7,7,4]$  &   8.26728345537e73 &  2.8e-4 &  3.15884774723e74 &  3.0e-4 &  3.82090163231 &  2.1e-5\\
$[7,6,8,7,5]$  &  8.26726695059e73 &  2.8e-4 &  3.15884117573e74 &  3.0e-4 &  3.82090139389 &  2.1e-5\\
$[7,7,7,7,5]$  &  8.26699151395e73 &  2.4e-4 &  3.15873118052e74 &  2.6e-4 &  3.82089567184 &  2.0e-5\\
$[7,8,6,7,5]$   &  8.26725194151e73 &  2.8e-4 &  3.15883519816e74 &  3.0e-4 &  3.82090115547 &  2.1e-5\\
$[7,6,7,7,6]$ &  8.26629485863e73 &  1.6e-4 &  3.15844126164e74 &  1.7e-4 &  3.82086682320 &  1.2e-5\\
$[7,7,6,7,6]$  &  8.26620018741e73 &  1.5e-4 &  3.15840205612e74 &  1.6e-4 &  3.82086324692 &  1.1e-5 \\
$[6,7,7,6,7]$  &  8.26745819354e73 &  3.0e-4 &  3.15891913390e74 &  3.2e-4 &  3.82090735435 &  2.3e-5\\
$[7,6,6,7,7]$  &  8.26695461959e73 &  2.4e-4 &  3.15871265798e74 &  2.6e-4 &  3.82089018822 &  1.8e-5\\
$[6,7,6,6,8]$  &  8.26663423405e73 &  2.0e-4 &  3.15857883123e74 &  2.2e-4 &  3.82087659836 &  1.5e-5\\

\hline

\end{tabular}
}
\end{center}
\label{dyck100}
\end{table}%
\section{Saving CPU time and detecting errors.}\label{sec:cpu}
In many enumeration problems where the coefficients are large integers, it is customary to perform the calculations {\em modulo} a large prime, repeat the calculation with different primes, and then reconstruct the coefficients using the Chinese Remainder Theorem. Using the idea of series extension that we have discussed, the initial digits of each coefficient can be predicted, and so fewer primes need to be used.

As an example, in recent unpublished work \cite{CG16} Conway and Guttmann extended their earlier work \cite{CG14} on the enumeration of 1324-avoiding permutations. The calculations were carried out modulo primes just smaller than $2^{62} \approx 4.6 \times 10^{18}.$ The series is known up to the coefficient of $z^{36}.$ We predicted from differential approximants the value of $c_{37} \approx 7.39 \times 10^{29} $ to 20 significant digits. That left only the last 9 or 10 digits uncertain. With one computer run, modulo a prime of the afore-mentioned size, we were able to establish the last 18 digits, and so could find the new coefficient exactly. Without this method, we would have needed to run the program again with a different prime. In this way we have halved the computer time required.

There are other situations in which even greater savings can be achieved.

Another useful application of the method is in detecting errors. It is sometimes the case that the last known coefficient in a series is in error, as the algorithm and computing resources are being pushed to their available limits. As we can often predict the first 15-20 digits of the first unknown coefficient of a series, using the available series without the last term has been, on occasion, sufficiently precise to suggest that the last known coefficient  is incorrect. The value of such a check is self-evident.

\section{Conclusion}
Given the first few coefficients of typical generating functions that arise in many problems of statistical mechanics or enumerative combinatorics, we have shown that the method of differential approximants performs surprisingly well in predicting subsequent coefficients. Perhaps even more surprisingly, this is also the case when the method of differential approximants does a poor job job in estimating the critical parameters, such as those cases in which one has stretched exponential behaviour. 

We have given examples that show how these extended series can be used as input to the ratio method to obtain significantly more precise estimates of the critical parameters. By predicting the most significant digits of unknown coefficients, this idea dovetails well with algorithms that predict the least significant digits of new coefficients by working modulo a prime. In this way, significant computer time can be saved. The predicted coefficients also provide a useful check on algorithms producing exact coefficients.

We believe that this method may open up a  new chapter in the method of series analysis.

\section{Acknowledgements}
I would like to thank Mireille Bousquet-M\'elou, Nathan Clisby and Jay Pantone for a critical reading of the manuscript, which resulted in substantial improvement.

\end{document}